\documentclass[12pt,oneside,a4paper]{book}

\usepackage{natbib}
\usepackage{times,amsmath,float}
\usepackage{amssymb}
\usepackage{graphicx}
\usepackage{amsfonts}
\usepackage{epsfig}
\usepackage{anysize}
\usepackage{xtab}
\usepackage{lscape}
\usepackage{geometry}
\usepackage{longtable}
\usepackage{appendix}
\usepackage{setspace}
\usepackage{epstopdf}
\usepackage{chngcntr}
\usepackage{booktabs}
\usepackage{multirow}
\usepackage[justification=justified,singlelinecheck=false,labelsep=period]{caption}
\usepackage{array}
\usepackage{adjustbox}
\usepackage{url}
\usepackage{multicol}
\usepackage{placeins}
\usepackage{pdfpages}
\usepackage{titling}
\usepackage{xcolor}
\usepackage{titlesec}
\usepackage{fancyhdr}
\usepackage{indentfirst}

\titleformat{\chapter}[display]
{\normalfont\normalsize\centering\upshape}
{\MakeUppercase{\chaptertitlename}\ \thechapter}
{8pt}
{\normalsize}

\titleformat{\section}
{\normalfont\normalsize\centering\upshape}
{}
{1em}
{}

\titleformat{\subsection}
{\normalfont\normalsize\centering\upshape}
{}
{1em}
{}

\titlespacing{\chapter}{0pt}{-50pt}{0.3cm}

\marginsize{4cm}{2.5cm}{2.5cm}{2.5cm}

\counterwithout{figure}{chapter}
\counterwithout{table}{chapter}
\counterwithout{footnote}{chapter}
\counterwithout{equation}{chapter}


\newcommand{\researchfigure}[2]{
\begin{figure}[H]
    \centering
    \IfFileExists{#1}{
        \includegraphics[width=0.90\textwidth]{#1}
    }{
        \fbox{
            \parbox[c][5cm][c]{0.85\textwidth}{
                \centering
                Upload figure file:\\
                \texttt{\detokenize{#1}}
            }
        }
    }
    \caption{#2}
\end{figure}
}

\usepackage{tocloft}

\begin{document}

\setlength{\parindent}{2.1em}
\doublespacing


\frontmatter


\begin{center}

\vspace{6cm}

\LARGE
Entrepreneurial Entry and Business Success:\\
The Roles of Gender, Nativity, Education, and Family Background

\vspace{4cm}

\Large Tuana Ozkan\\
\&\\
\Large Bayram Cakir

\vspace{4cm}

\Large November 2026

\end{center}

\pagebreak


\chapter{ABSTRACT}
\vspace*{0.4cm}

\indent This study examines entrepreneurship as a two-stage process: entry into business ownership and post-entry business outcomes. Using data from the National Longitudinal Survey of Youth 1979 (NLSY79), the analysis compares unweighted descriptive rates across sex, nativity status, individual educational attainment, parental educational attainment, and early-life family income. Nativity is defined by country of birth, and all reported relationships are interpreted as descriptive associations rather than causal effects. \\

\indent The business-ownership analysis includes 8,201 respondents with a valid response to whether they had ever owned a qualifying business, of whom 1,231 reported having done so. The clearest entry pattern is by sex: 18.2\% of male respondents (\(n=4{,}028\)) report business ownership, compared with 11.9\% of female respondents (\(n=4{,}173\)). The male ownership rate remains higher within every educational, parental-education, and family-income category examined, although the gap becomes very small in the doctoral/professional group. Ownership also tends to be higher in several more-educated and higher-family-income groups, but these relationships are not monotonic. By contrast, U.S.-born and foreign-born respondents have relatively similar overall ownership rates, 15.1\% (\(n=7{,}455\)) and 13.9\% (\(n=562\)), and the direction of the nativity difference varies across subgroups. \\

\indent The post-entry analysis is restricted to respondents who established Business 01 themselves or with partners. A business-success proxy classifies continued ownership or sale as a successful outcome and shutdown as an unsuccessful outcome. Among 629 founders with a valid outcome, 36.7\% are classified as successful under this definition. Male and female founders have nearly identical aggregate proxy-success rates, 36.9\% (\(n=371\)) and 36.4\% (\(n=258\)), respectively. Foreign-born founders have a higher observed rate than U.S.-born founders, 45.9\% (\(n=37\)) versus 36.3\% (\(n=576\)), but the foreign-born sample is too small to support stable subgroup conclusions. Education, parental education, and family income show some differences in post-entry outcomes, but these patterns are less consistent than the entry patterns and many estimates are based on small founder samples. \\

\indent Overall, the most robust descriptive finding is a strong gender difference in entrepreneurial entry that is not mirrored by the aggregate business-success proxy among observed founders. This contrast highlights the value of separating who enters entrepreneurship from what is observed after entry, while the study's descriptive design, unweighted estimates, outcome proxy, and small founder subgroups limit causal or population-level conclusions.

\pagebreak


\clearpage

\begingroup

\setstretch{1.35}

\tableofcontents

\endgroup

\doublespacing

\clearpage


\mainmatter


\chapter{INTRODUCTION}
\vspace*{0.4cm}
\doublespacing

\indent Entrepreneurship is commonly associated with innovation, employment creation, economic mobility, and individual independence. However, access to entrepreneurship is not distributed equally among individuals. Starting a business may depend not only on entrepreneurial ability or motivation, but also on access to financial resources, education, experience, family history, and social networks. These differences may influence both whether an individual becomes a business owner and what happens after a business has been established. \\

\indent One of the most persistent differences observed in entrepreneurship is the gender gap. Previous research has shown that women are less likely than men to own businesses and that female-owned businesses can differ from male-owned businesses in survival, sales, employment, and access to capital. Fairlie and Robb (2009), for example, document differences in business performance between female and male-owned firms and identify startup capital and previous business experience as important factors. However, the existence of a gender gap in business ownership does not necessarily imply that an equally large difference will remain among individuals who successfully enter entrepreneurial activities. \\

\indent Education and family background may also influence entrepreneurship. Human-capital theories suggest that education, managerial experience, and a broad range of skills can increase an individual's ability to identify and exploit business opportunities. Kim, Aldrich, and Keister (2006) find that advanced education and managerial experience are positively related to entrepreneurial entry. Lazear (2005) similarly argued that entrepreneurs benefit from possessing a balanced range of skills rather than specializing narrowly in one area. Education may therefore influence both entry into business ownership and the ability to maintain a business after entry. \\

\indent Financial and family resources provide another potential source of differences in entrepreneurship. Evans and Jovanovic (1989) argued that liquidity constraints may prevent individuals with insufficient capital from entering entrepreneurship. However, the relationship between wealth and entrepreneurship is not necessarily linear. Hurst and Lusardi (2004) find that the probability of business entry remains relatively flat across much of the wealth distribution and increases substantially only at very high levels of wealth. Family background may also matter through nonfinancial mechanisms. Dunn and Holtz-Eakin (2000), for example, identified strong intergenerational relationships in self-employment and emphasize parental entrepreneurial experience and human capital. \\

\indent Nativity status provides another important dimension. Immigrants contribute substantially to the formation of companies in the United States. Kerr and Kerr (2020) estimate that first-generation immigrants account for approximately one-quarter of newly created firms in the United States. However, immigrant and native-owned businesses can differ in industry, geography, networks, resources, and other characteristics. It is therefore useful to distinguish between entrepreneurial entry and business outcomes rather than assuming that differences in business formation necessarily correspond to the same differences after entry. \\

\indent This study brings these dimensions together using the National Longitudinal Survey of Youth 1979 (NLSY79). The NLSY79 is particularly useful because it follows the same individuals over a long period and combines detailed information on demographic characteristics, family background, education, income, employment, and business ownership history. \\

\indent The first question addressed in this study is how sex, nativity status, educational attainment, parental educational attainment, and family income are associated with business ownership. The second question is whether these same characteristics are associated with business outcomes among respondents who have established businesses themselves or with partners. \\

\indent An important aspect of the analysis is therefore the distinction between entering entrepreneurship and experiencing success after entry. The results show a strong and persistent gender gap in business ownership, but almost no overall gender difference in the business-success measure between founders. Nativity status produces a different pattern: U.S.-born and foreign-born respondents have relatively similar overall ownership rates, while foreign-born founders have a higher observed success rate. However, the relatively small number of foreign-born founders limits the conclusions that can be drawn from the latter result. \\

\indent The remainder of the paper is organized as follows. Chapter 2 reviews the existing literature on gender, human capital, family resources, financial constraints, and nativity in entrepreneurship. Chapter 3 defines the data and variables used in the study. Chapter 4 presents the business ownership analysis. Chapter 5 analyzes business success among founders. Chapter 6 presents the conclusion and discusses the limitations of the analysis.

\clearpage


\chapter{LITERATURE REVIEW}
\vspace*{0.4cm}
\doublespacing

\indent Research on entrepreneurship has long attempted to explain why some individuals enter business ownership while others remain in paid employment. A related body of research examines why some businesses survive or expand while others close. Although these questions are connected, the determinants of entrepreneurial entry and post-entry outcomes do not necessarily have to be identical. \\

\indent Gender is one of the most extensively studied dimensions of entrepreneurial inequality. Fairlie and Robb (2009) analyze U.S. Census Bureau business-owner data and find differences between female-owned and male-owned businesses in survival, profits, employment, and sales. Their analysis suggests that differences in startup capital, prior work experience in a similar business, and experience in family businesses contribute to these performance differences. \\

\indent These findings suggest that gender differences in entrepreneurship can operate at two different stages. The first occurs during entrepreneurial entry, when women may be less likely than men to start or acquire businesses. The second occurs after entry, when business outcomes may differ among individuals who have already entered entrepreneurship. Separating these stages is therefore important. \\

\indent Human capital represents another important explanation for entrepreneurial behavior. Kim, Aldrich, and Keister (2006) examine financial, human, and cultural capital in entrepreneurial entry and find that advanced education and managerial experience are positively associated with startup activity. Their findings suggest that skills and experience can play an important role in an individual's decision and ability to enter entrepreneurship. \\

\indent Lazear (2005) provides a complementary explanation through the ``jack-of-all-trades'' theory of entrepreneurship. According to this theory, successful entrepreneurs require competence across several different areas rather than exceptional specialization in only one skill. Using data on Stanford alumni, Lazear finds that individuals with broader educational and professional backgrounds are more likely to become entrepreneurs. \\

\indent Education may also influence outcomes after entry. Bates (1990), examining small-business longevity, identifies educational attainment as an important human-capital characteristic related to whether businesses remain in operation. Higher education may improve managerial ability, financial decision-making, opportunity recognition, or access to external resources. For this reason, educational attainment is considered in both stages of the analysis in this study. \\

\indent Financial resources have also received substantial attention in the entrepreneurship literature. Evans and Jovanovic (1989) provide evidence consistent with liquidity constraints and argue that individuals who lack sufficient financial resources can be prevented from entering entrepreneurship even when they possess entrepreneurial ability. Under this explanation, individuals from financially stronger backgrounds may have greater opportunities to become business owners. \\

\indent The empirical relationship between financial resources and entrepreneurship, however, is not simple. Hurst and Lusardi (2004) find that the relationship between household wealth and business entry is relatively flat across most of the wealth distribution. A strong positive relationship emerges primarily among households near the top of the distribution. Their findings therefore suggest that the effect of financial resources on entrepreneurship may be nonlinear. \\

\indent Family background can influence entrepreneurship through channels other than direct financial support. Dunn and Holtz-Eakin (2000) use National Longitudinal Survey data to study transitions into self-employment and find that parental factors have a substantial influence. In particular, parental self-employment experience and business success appear to be important mechanisms through which entrepreneurial characteristics are transmitted across generations. \\

\indent Hout and Rosen (2000) similarly find strong intergenerational relationships in self-employment. Individuals whose fathers were self-employed are more likely to become self-employed themselves. This evidence suggests that exposure to entrepreneurial knowledge, networks, experience, and expectations within the family may affect future business ownership. \\

\indent These findings motivate the inclusion of parental educational attainment in the present study. Although parental education is not equivalent to parental entrepreneurial experience, it provides information about the human-capital and socioeconomic environment in which the respondent grew up. Family income provides a complementary measure of early-life financial background. \\

\indent Nativity status is another relevant characteristic in the entrepreneurship literature. Kerr and Kerr (2020), using the U.S. Survey of Business Owners, find that immigrant entrepreneurs create approximately one-quarter of new firms in the United States. Immigrant entrepreneurship may be influenced by self-selection into migration, labor-market barriers, ethnic and international networks, access to specific opportunities, and differences in human capital. At the same time, immigrants may face constraints in credit markets, language, information, or professional networks. Therefore, the expected relationship between nativity and entrepreneurship is theoretically ambiguous. \\

\indent Existing research provides strong reasons to expect gender, education, family resources, and nativity to be associated with entrepreneurship. However, it also suggests that their relationships may differ between entrepreneurial entry and business outcomes. The present study therefore treats business ownership and business success as related but separate dependent variables.

\clearpage


\chapter{THE DATA}
\vspace*{0.4cm}
\doublespacing

\indent The data used in this study are obtained from the National Longitudinal Survey of Youth 1979 (NLSY79), which is administered as part of the National Longitudinal Surveys program of the U.S. Bureau of Labor Statistics. The original NLSY79 cohort consisted of 12,686 respondents who were first interviewed in 1979. The respondents were born between 1957 and 1964 and were approximately 14 to 22 years old when the first interviews were conducted. \\

\indent The longitudinal structure of the NLSY79 allows information collected at different stages of respondents' lives to be combined. The survey contains extensive information on education, employment, income, demographic characteristics, family background, and other socioeconomic outcomes. Beginning in the 2010 survey round, current and former business owners were asked a detailed series of questions about businesses they had owned since age 18. Cross-round business ownership variables integrate information collected from 2010 onward. \\

\indent The NLSY79 definition of business ownership includes businesses operated regularly for the purpose of generating income or profit. It includes incorporated companies, partnerships, sole proprietorships, independent contracting, consulting, freelancing, and qualifying informal businesses. Activities operated only sporadically, activities conducted purely as hobbies, and passive investments without an operational role are excluded. \\

\indent The primary dependent variable in the first part of the analysis is \textit{Ever Owned a Business}. The variable equals 1 if a respondent reports having owned a qualifying business since age 18 and equals 0 if the respondent reports not having owned such a business. \\

\indent The full dataset contains 12,686 respondents. Of these respondents, 8,201 have a valid 0 or 1 response for the business ownership variable. Among these valid respondents, 1,231 report having owned a business, while 6,970 report that they have not. The overall unweighted ownership rate in the analytical sample is therefore approximately 15.0\%. \\

\indent For all business ownership subgroup analyses, the reported sample size represents the number of respondents who have a valid 0 or 1 ownership response and valid information for the subgroup characteristic being examined. Both business owners and respondents who have never owned a business are therefore included in the denominator. Missing business ownership observations are excluded. \\

\indent The first demographic characteristic used in the analysis is sex. The original NLSY79 coding distinguishes between male and female respondents. In the cleaned dataset, these values are labeled \textit{Male} and \textit{Female}. \\

\indent Nativity status is constructed from the respondent's country of birth. Respondents born in the United States are categorized as \textit{U.S.-born}, while respondents born outside the United States are categorized as \textit{Foreign-born}. Therefore, the foreign-born category represents birthplace rather than legal immigration status. \\

\indent The respondent's own educational attainment is measured using the highest academic degree ever received. The original categories are combined into five broader groups: high school, associate degree, bachelor's degree, master's degree, and doctoral/professional degree. Bachelor of Arts and Bachelor of Science degrees are combined in the bachelor's category, while doctoral and professional degrees are combined in the doctoral/professional category. \\

\indent Mother's and father's educational attainment are based on the highest grade or year of schooling completed by each parent. Values below 12 years are categorized as \textit{Less than high school}; 12 years is categorized as \textit{High school}; 13--15 years as \textit{Some college}; 16 years as \textit{College graduate}; and 17--20 years as \textit{Postgraduate}. \\

\indent Family income is measured using total net family income reported in the 1979 survey. The income measure represents family income during the previous calendar year, 1978. It is therefore interpreted as an indicator of early-life family socioeconomic background rather than the respondent's financial position at the exact time of business formation. \\

\indent Valid family-income observations are divided into four groups according to quartiles of the analytical sample. The 25th percentile is \$7,000, the median is \$13,050, and the 75th percentile is \$23,000. Respondents are therefore categorized into low, lower-middle, upper-middle, and high family-income groups. These categories are relative to the sample and should not be interpreted as official U.S. income classifications. \\

\indent The original NLSY79 data contain negative numerical codes representing missing or non-substantive responses. These codes include refusal, don't know, invalid skip, valid skip or question not applicable, and non-interview. These values are treated as missing observations in the cleaned dataset instead of being interpreted as substantive numerical responses. Consequently, the number of observations differs across analyses depending on the availability of valid information. \\

\indent The second part of the study focuses on business success among respondents who founded a business. The NLSY79 records how ownership of Business 01 was acquired. To focus specifically on founders, the success analysis is restricted to respondents who established the business either alone or with partners. Respondents who obtained ownership through a gift, purchase, inheritance, marriage, or another mechanism are excluded from the founder-success analysis. \\

\indent There are 683 Business 01 founders in the dataset. A binary business-success proxy is constructed for these founders. A founder is coded as successful if the respondent still maintains ownership of Business 01 or reports that ownership ended because the business was sold. A founder is coded as unsuccessful if the business was shut down. Respondents whose ownership ended for ``other'' reasons, or for whom a clear outcome cannot be determined, are treated as missing. \\

\indent This definition results in 629 founders with valid business-success outcomes. Among them, 231 are classified as successful and 398 as unsuccessful, producing an overall success rate of approximately 36.7\%. \\

\indent The business-success measure should be interpreted as a proxy rather than as a direct measure of profitability. Continued ownership does not necessarily indicate that a business is profitable, while the sale of a business does not necessarily indicate a profitable exit. The measure instead distinguishes businesses that remain owned or were sold from those explicitly reported as having shut down. \\

\indent The analysis in this study is descriptive. Business ownership and business-success proxy rates are compared across demographic and socioeconomic subgroups. No regression adjustment or causal identification strategy is used, and no causal relationship is inferred from these descriptive comparisons. The number of valid observations underlying each percentage is also considered when interpreting the results. This is especially important in the business-success analysis, where some subgroups contain very small numbers of founders. \\

\indent The statistics reported are unweighted descriptive statistics from the analytical sample. Since the original NLSY79 used a complex sampling design that included supplemental samples, the percentages reported in this paper should not automatically be interpreted as population-weighted estimates for the entire United States. \\

\indent A final limitation concerns timing. The respondent's highest educational degree is a lifetime measure, while a respondent may have founded a business before completing that degree. Therefore, educational attainment is interpreted as a characteristic associated with business ownership and success rather than as a clearly timed causal treatment.

\clearpage


\chapter{BUSINESS OWNERSHIP ANALYSIS}
\vspace*{0.4cm}
\doublespacing

\indent This chapter examines business-ownership rates across sex and nativity status, then uses individual education, parental education, and early-life family income to determine whether the same broad patterns appear within different background groups. The comparisons are descriptive, and the reported \(n\) values are the valid denominators underlying each rate. \\


\section{Gender Differences in Business Ownership}

\indent The aggregate ownership comparison shows a clear gender difference. The ownership rate is 18.2\% among male respondents (\(n=4{,}028\)) and 11.9\% among female respondents (\(n=4{,}173\)), a difference of approximately 6.4 percentage points. In relative terms, the male rate is about 1.5 times the female rate. \\

\researchfigure{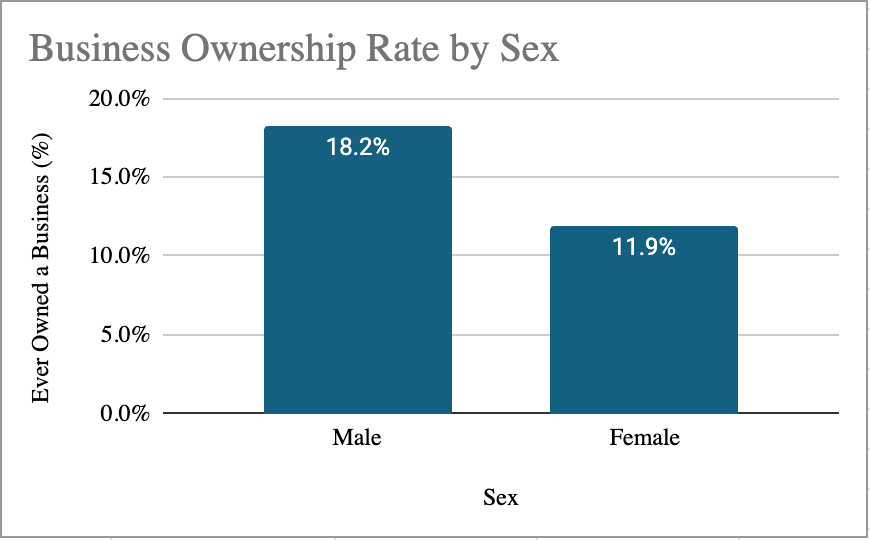}
{Business Ownership Rate by Sex}

\indent The gender gap remains visible across educational categories. Among high-school graduates, ownership is 16.4\% for men (\(n=2{,}309\)) and 11.3\% for women (\(n=2{,}160\)). Among associate-degree holders, the rates are 20.4\% (\(n=397\)) and 12.3\% (\(n=594\)); among bachelor's-degree holders, 25.5\% (\(n=608\)) and 15.2\% (\(n=664\)); and among master's-degree holders, 19.7\% (\(n=208\)) and 12.4\% (\(n=346\)). \\

\indent The doctoral/professional category is the one educational group in which the gender gap almost disappears: 28.0\% of men (\(n=93\)) and 27.8\% of women (\(n=54\)) report business ownership. Because these denominators are much smaller than those in the high-school and bachelor's groups, the near equality should be interpreted cautiously. \\

\researchfigure{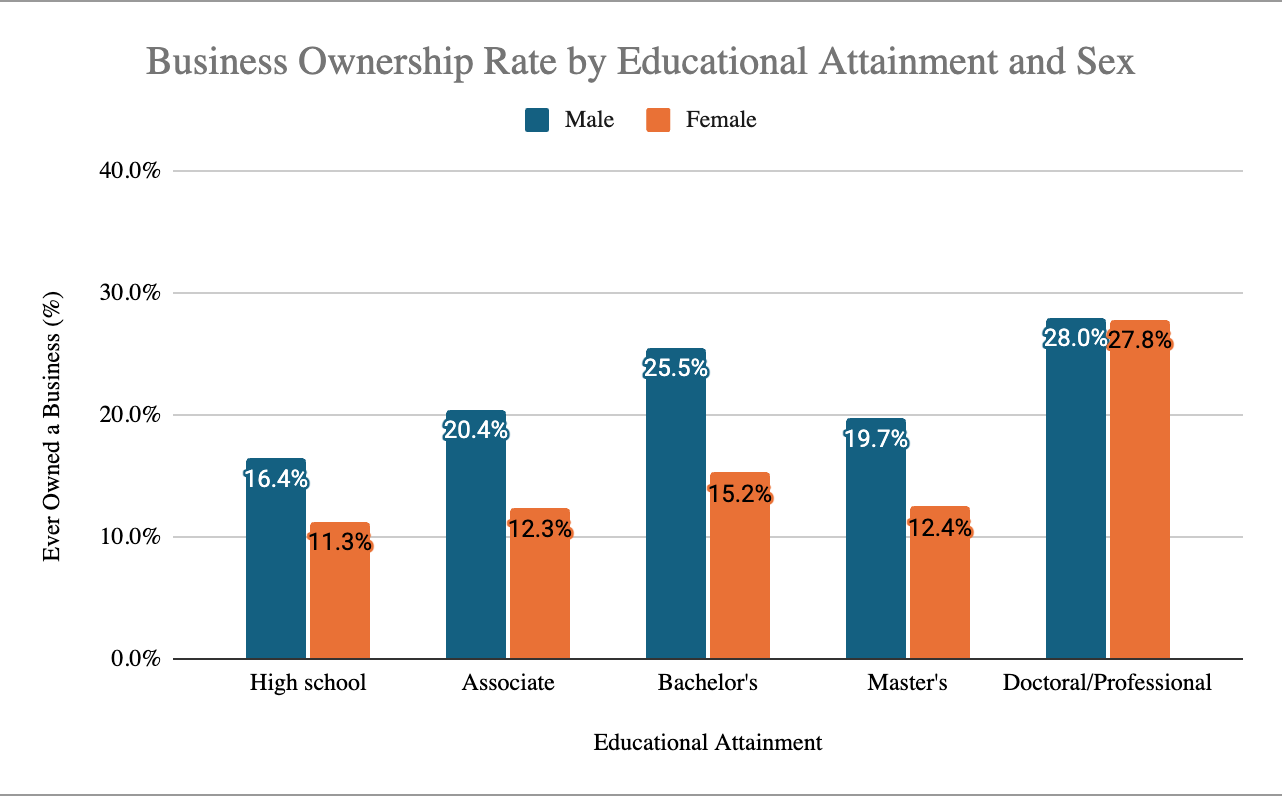}
{Business Ownership Rate by Educational Attainment and Sex}

\indent Two features stand out in Figure 2. First, ownership tends to be higher in several of the more advanced educational categories than among high-school graduates, but the pattern is not monotonic: for both sexes, the rate falls at the master's level relative to the bachelor's level before rising in the doctoral/professional group. Second, the male rate is higher in every educational category, although the difference is only 0.2 percentage points in the doctoral/professional group. Education therefore stratifies ownership rates, but the descriptive gender difference remains within each of these broad degree categories. \\

\indent Maternal education shows a broadly increasing ownership pattern. When the mother completed less than high school, ownership is 14.6\% for men (\(n=1{,}560\)) and 8.9\% for women (\(n=1{,}815\)). When the mother completed high school, the rates are 19.9\% (\(n=1{,}535\)) and 13.4\% (\(n=1{,}489\)); with some college, 21.0\% (\(n=352\)) and 15.6\% (\(n=360\)); with a college degree, 28.2\% (\(n=209\)) and 21.6\% (\(n=208\)); and with postgraduate education, 29.1\% (\(n=79\)) and 20.9\% (\(n=67\)). \\

\researchfigure{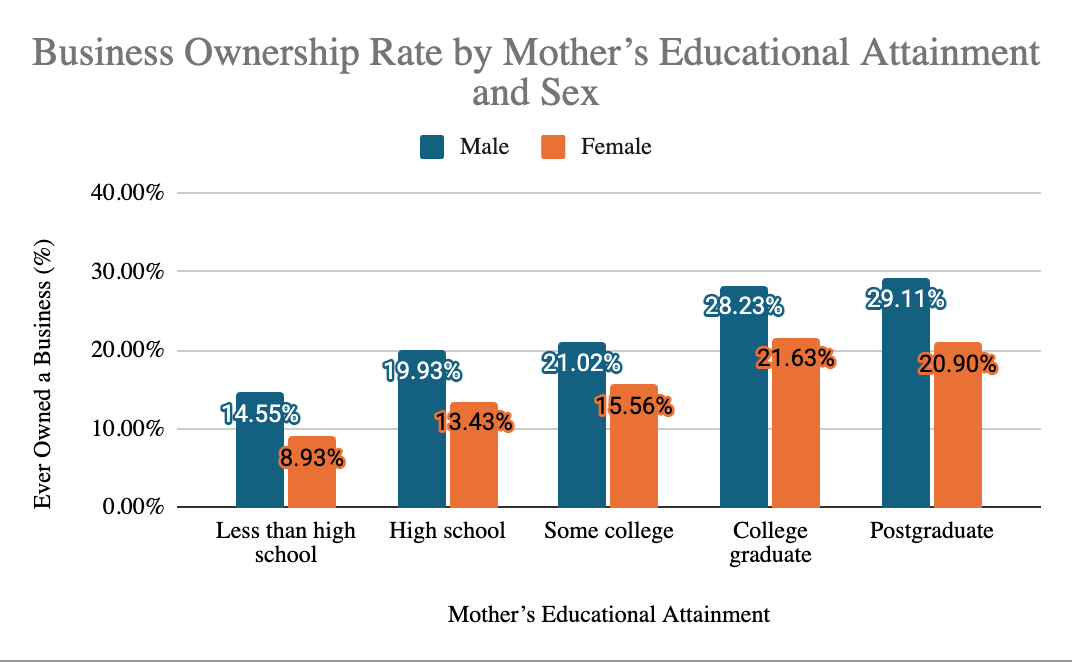}
{Business Ownership Rate by Mother's Educational Attainment and Sex}

\indent The maternal-education pattern is comparatively consistent. Ownership generally rises as maternal education increases, especially when the college-graduate and postgraduate groups are compared with the less-than-high-school group. At the same time, the male ownership rate remains above the female rate in all five categories. The smallest denominators occur in the postgraduate group, so the exact size of that final difference is less precise than the pattern in the larger categories. \\

\indent Paternal education produces a similar but slightly less monotonic pattern. When the father completed less than high school, ownership is 14.3\% for men (\(n=1{,}421\)) and 8.9\% for women (\(n=1{,}558\)). The rates are 21.0\% (\(n=1{,}167\)) and 13.3\% (\(n=1{,}203\)) when the father completed high school; 21.6\% (\(n=338\)) and 15.3\% (\(n=313\)) with some college; 26.1\% (\(n=291\)) and 18.1\% (\(n=276\)) for college graduates; and 23.8\% (\(n=206\)) and 20.6\% (\(n=180\)) for postgraduate education. \\

\researchfigure{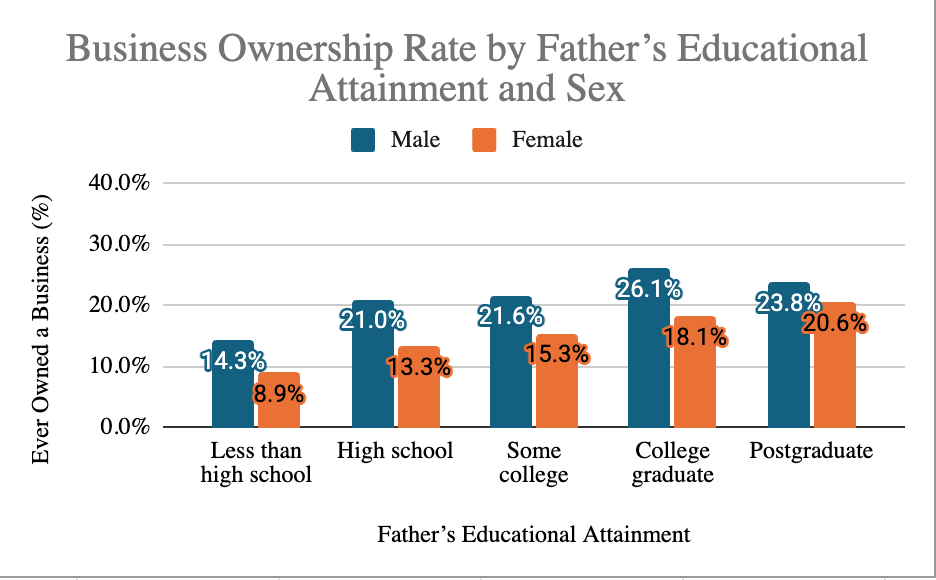}
{Business Ownership Rate by Father's Educational Attainment and Sex}

\indent Ownership is therefore broadly higher at higher levels of paternal education, but the relationship is not monotonic because both male and female rates change at the postgraduate level. The male rate nevertheless remains higher in every paternal-education category. Compared with Figure 3, Figure 4 provides weaker evidence of a smooth education gradient but equally clear evidence that the aggregate gender ordering persists within these broad groups. \\

\indent Early-life family income produces another visible pattern. In the low-income group, ownership is 15.6\% for men (\(n=831\)) and 9.3\% for women (\(n=906\)); in the lower-middle group, 14.7\% (\(n=788\)) and 8.9\% (\(n=783\)); in the upper-middle group, 20.4\% (\(n=829\)) and 13.4\% (\(n=866\)); and in the high-income group, 21.4\% (\(n=846\)) and 16.4\% (\(n=764\)). \\

\researchfigure{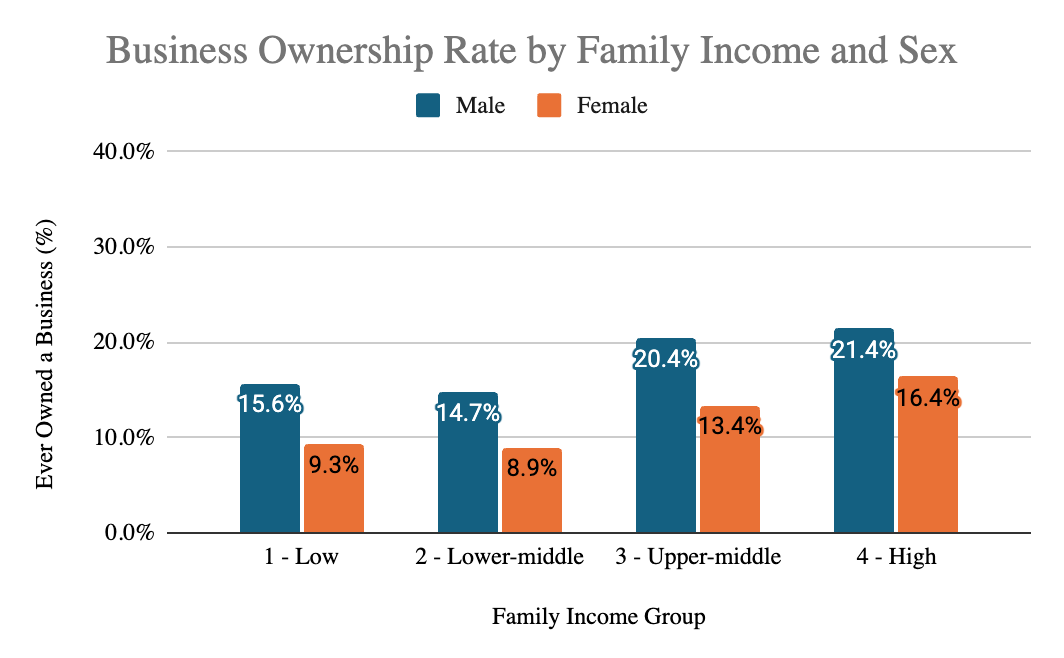}
{Business Ownership Rate by Family Income and Sex}

\indent The low and lower-middle groups are very similar within each sex, while ownership rises more clearly in the upper half of the family-income distribution. This creates a threshold-like descriptive pattern rather than a steady linear increase from one quartile to the next. The male rate remains higher in all four income groups, with gaps ranging from approximately 5.0 to 7.0 percentage points. \\

\indent Taken together, Figures 1--5 show that the gender ownership gap is not confined to a single education or family-background category. The male rate is higher overall and within every subgroup shown. Because these are separate descriptive comparisons rather than a multivariable model, they do not establish that education or family background ``fails to explain'' the gender gap in a causal or statistical-adjustment sense. They do show, however, that the aggregate gap is not driven only by one unusual subgroup and remains visible when men and women are compared within these measured background categories. \\

\indent The background variables themselves also matter descriptively. Ownership is often higher among respondents with more education, more highly educated parents, or family income in the upper half of the sample, but none of these relationships is perfectly linear. The strongest conclusion from this section is therefore the consistency of the gender ordering rather than a claim that any one background variable has a uniform effect on ownership.


\section{Business Ownership and Nativity Status}

\indent The aggregate nativity comparison is much smaller than the gender comparison. The ownership rate is 15.1\% among U.S.-born respondents (\(n=7{,}455\)) and 13.9\% among foreign-born respondents (\(n=562\)), a difference of approximately 1.2 percentage points. \\

\researchfigure{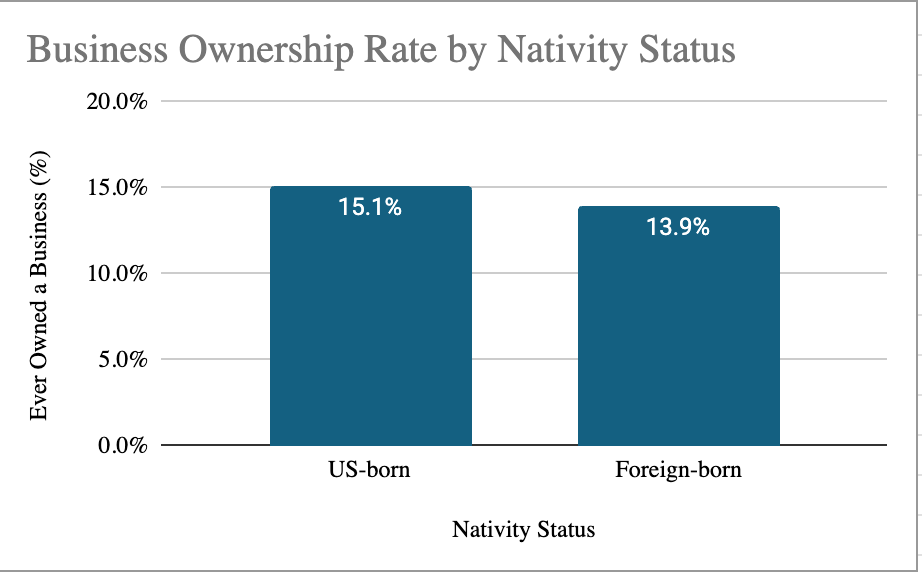}
{Business Ownership Rate by Nativity Status}

\indent The direction of the nativity difference changes across education groups. Among high-school graduates, ownership is 14.0\% for U.S.-born respondents (\(n=4{,}076\)) and 12.6\% for foreign-born respondents (\(n=278\)). Among associate-degree holders, the foreign-born rate is slightly higher, 17.2\% (\(n=64\)) versus 15.2\% (\(n=914\)); the same is true among bachelor's-degree holders, 22.2\% (\(n=63\)) versus 20.0\% (\(n=1{,}204\)). \\

\indent At the master's level, U.S.-born respondents have a higher rate, 15.6\% (\(n=519\)) compared with 9.1\% (\(n=33\)) among foreign-born respondents. The largest apparent difference occurs in the doctoral/professional group, 29.4\% for U.S.-born respondents (\(n=136\)) and 9.1\% for foreign-born respondents (\(n=11\)). The very small foreign-born denominators at the master's and doctoral/professional levels mean that those large percentage differences are particularly unstable. \\

\researchfigure{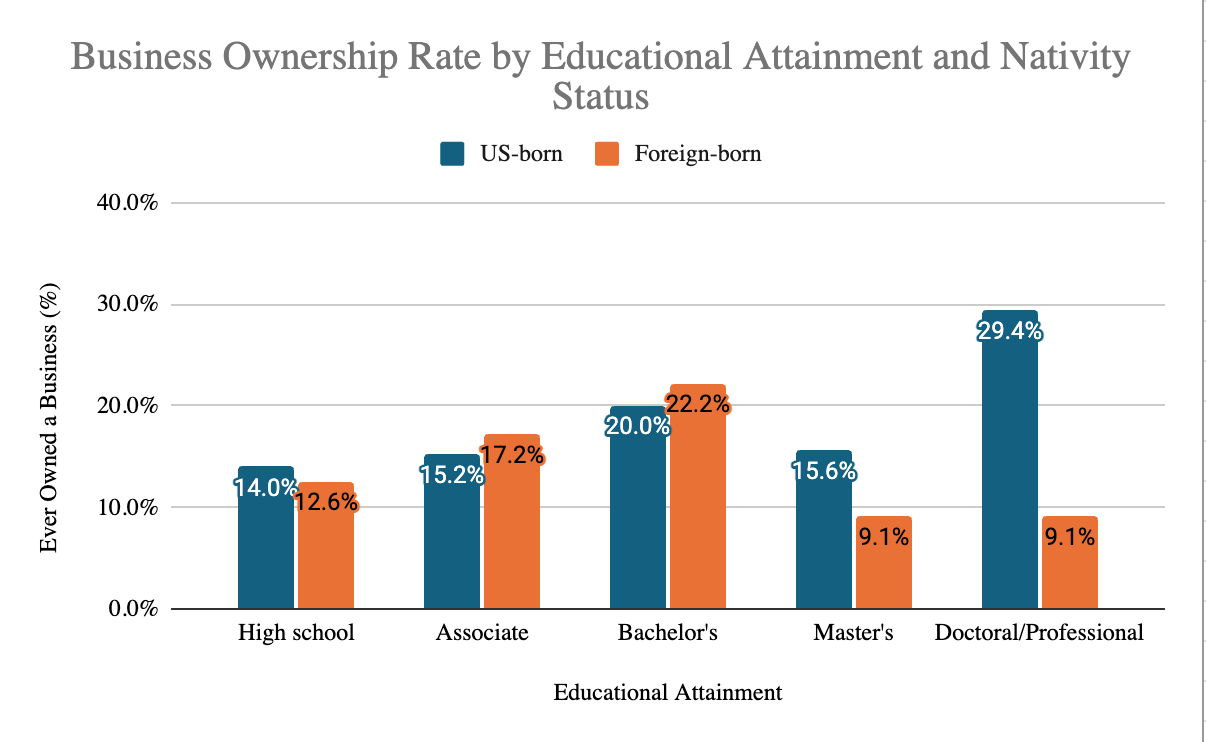}
{Business Ownership Rate by Educational Attainment and Nativity Status}

\indent Figure 7 therefore does not show a consistent nativity advantage across educational attainment. Foreign-born rates are slightly higher in the associate and bachelor's groups, while U.S.-born rates are higher in the high-school, master's, and doctoral/professional groups. The instability of the smallest foreign-born cells makes the higher-degree comparisons descriptive rather than conclusive. \\

\indent Maternal education produces the same lack of a consistent nativity ordering. When the mother completed less than high school, ownership is 11.1\% among U.S.-born respondents (\(n=2{,}947\)) and 13.9\% among foreign-born respondents (\(n=338\)). With a high-school-educated mother, the rates are 17.1\% (\(n=2{,}849\)) and 9.9\% (\(n=121\)); with some college, 18.0\% (\(n=683\)) and 20.0\% (\(n=20\)); with a college degree, 24.8\% (\(n=391\)) and 27.3\% (\(n=22\)); and with postgraduate education, 25.2\% (\(n=139\)) and 33.3\% (\(n=6\)). \\

\researchfigure{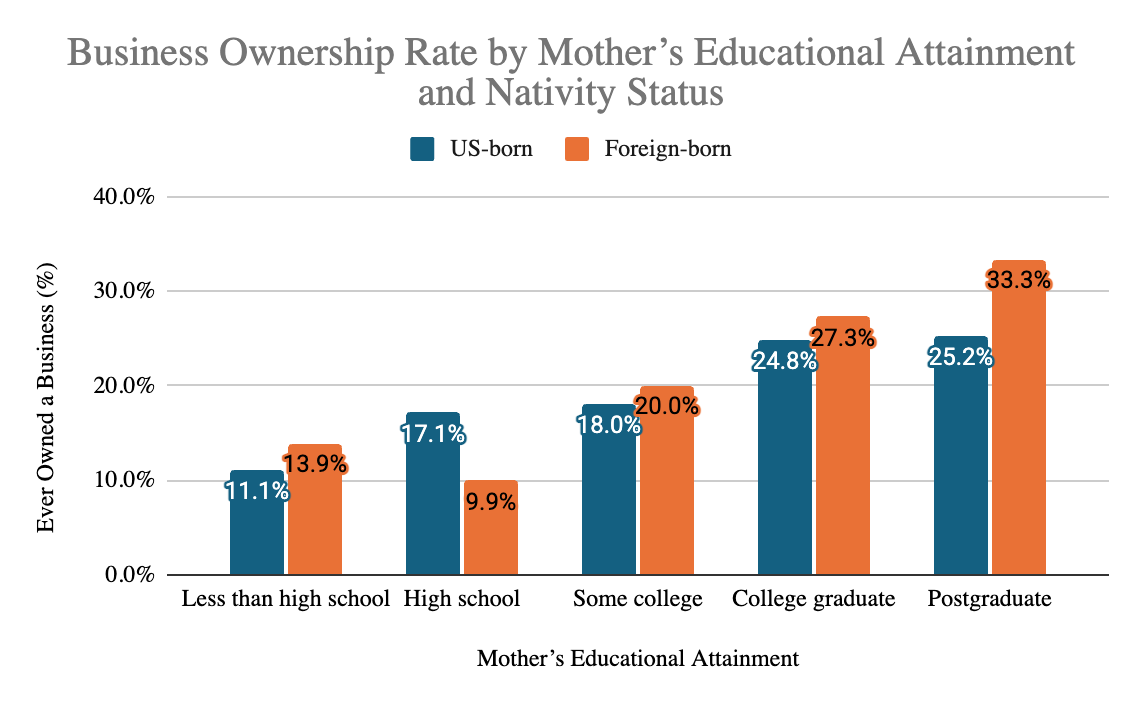}
{Business Ownership Rate by Mother's Educational Attainment and Nativity Status}

\indent The apparent foreign-born advantage in the highest maternal-education categories rests on only 20, 22, and 6 foreign-born respondents in the some-college, college-graduate, and postgraduate groups. Those percentages can move substantially with only one or two observations. The U.S.-born series shows a clearer broad increase with maternal education, while the foreign-born series is too sparse at higher levels for a stable trend. \\

\indent Paternal education again produces mixed nativity differences. When the father completed less than high school, ownership is 11.3\% for U.S.-born respondents (\(n=2{,}610\)) and 11.4\% for foreign-born respondents (\(n=281\)). With a high-school-educated father, the rates are 17.1\% (\(n=2{,}222\)) and 17.8\% (\(n=118\)); with some college, 19.0\% (\(n=617\)) and 11.5\% (\(n=26\)); with a college degree, 22.3\% (\(n=530\)) and 20.0\% (\(n=30\)); and with postgraduate education, 22.4\% (\(n=361\)) and 19.0\% (\(n=21\)). \\

\researchfigure{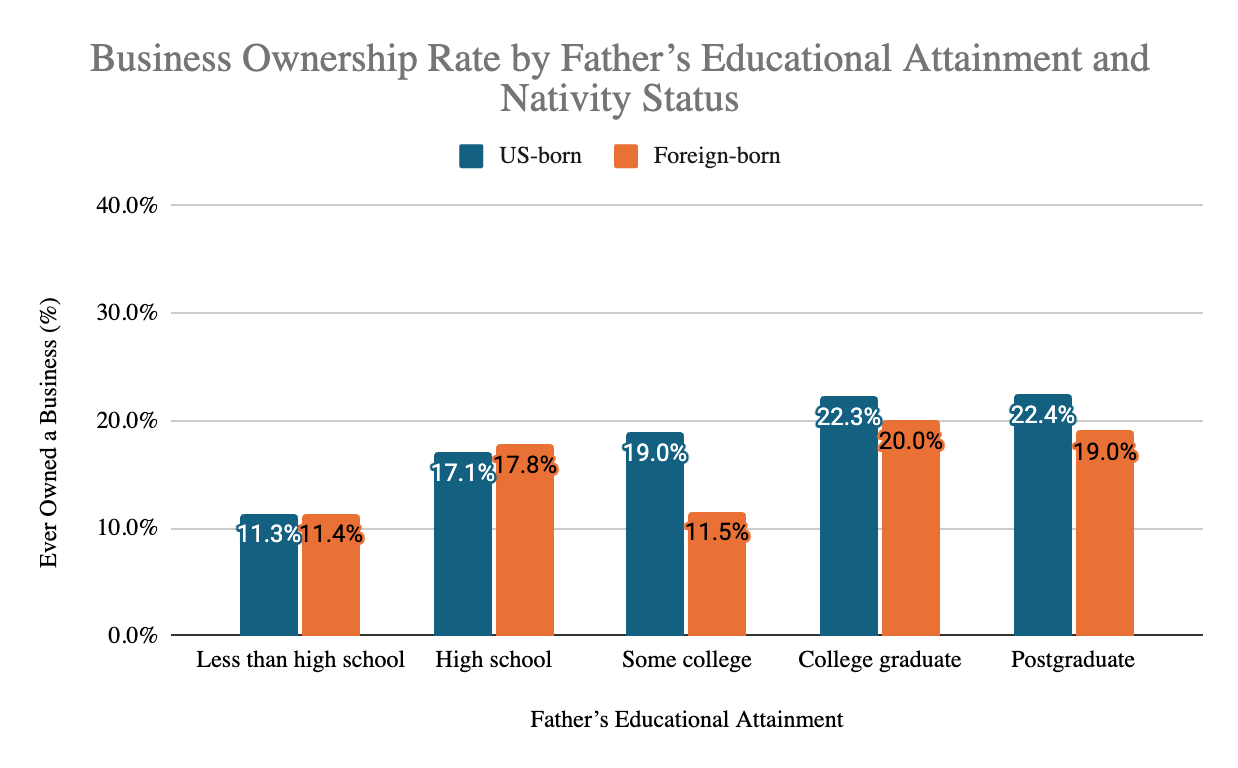}
{Business Ownership Rate by Father's Educational Attainment and Nativity Status}

\indent The two nativity groups are nearly identical in the two lowest paternal-education categories. U.S.-born respondents have higher observed rates in the three higher categories, but the foreign-born denominators fall to 26, 30, and 21. The direction of the difference is therefore more consistent here than in the maternal-education figure, but the high-education foreign-born estimates remain imprecise. \\

\indent Family income likewise does not produce a stable nativity ordering. In the low-income group, ownership is 12.3\% for U.S.-born respondents (\(n=1{,}568\)) and 14.2\% for foreign-born respondents (\(n=113\)); in the lower-middle group, 11.5\% (\(n=1{,}382\)) and 10.6\% (\(n=142\)); in the upper-middle group, 17.1\% (\(n=1{,}553\)) and 11.8\% (\(n=119\)); and in the high-income group, 19.0\% (\(n=1{,}521\)) and 20.0\% (\(n=70\)). \\

\researchfigure{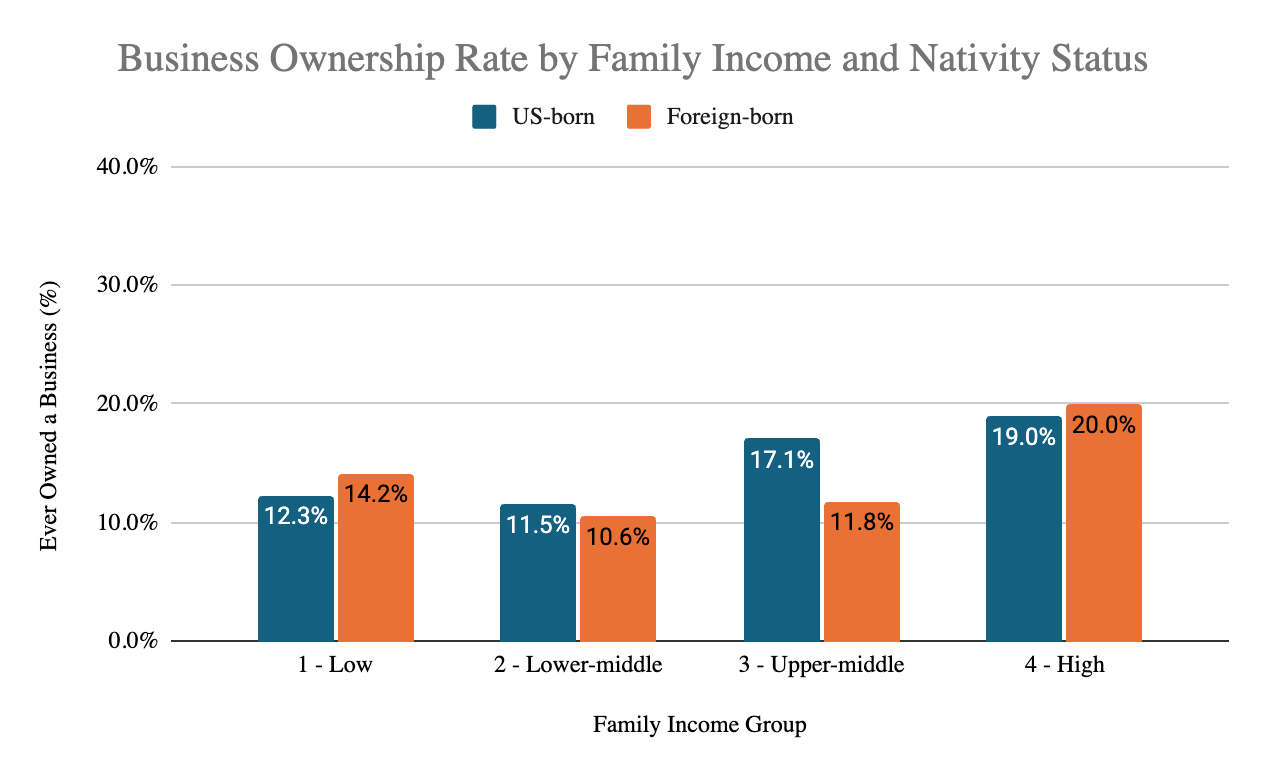}
{Business Ownership Rate by Family Income and Nativity Status}

\indent For U.S.-born respondents, ownership is clearly higher in the upper two income groups than in the lower two. The foreign-born series is less regular: the high-income group has the highest rate, but the low-income rate exceeds both middle groups. Most importantly for the nativity comparison, the higher rate alternates between U.S.-born and foreign-born respondents across quartiles. \\

\indent Overall, Figures 6--10 do not show a large, consistently directed nativity gap in business ownership. The aggregate difference is small, the ordering changes across several background categories, and the foreign-born sample becomes especially small at high levels of education. The data therefore support a stronger conclusion about gender differences in entry than about nativity differences in entry.

\clearpage


\chapter{BUSINESS SUCCESS ANALYSIS}
\vspace*{0.4cm}
\doublespacing

\indent The previous chapter examined entry into business ownership. This chapter considers a different outcome among respondents who established Business 01 themselves or with partners and have a valid observed outcome. The business-success measure is the proxy defined in Chapter 3: continued ownership or sale is coded as successful, while shutdown is coded as unsuccessful. It should therefore be interpreted as a limited post-entry outcome measure rather than as direct evidence of profitability or firm quality. \\

\indent The analytical sample contains 629 founders with valid proxy outcomes. Because this is much smaller than the ownership sample, subgroup comparisons often rely on modest or very small denominators. The interpretation below therefore gives particular weight to sample size and avoids treating extreme percentages from small cells as stable estimates. \\


\section{Gender Differences in Business Success}

\indent The aggregate gender result differs sharply from the ownership result. The proxy-success rate is 36.9\% among male founders (\(n=371\)) and 36.4\% among female founders (\(n=258\)), a difference of only about 0.5 percentage points. \\

\researchfigure{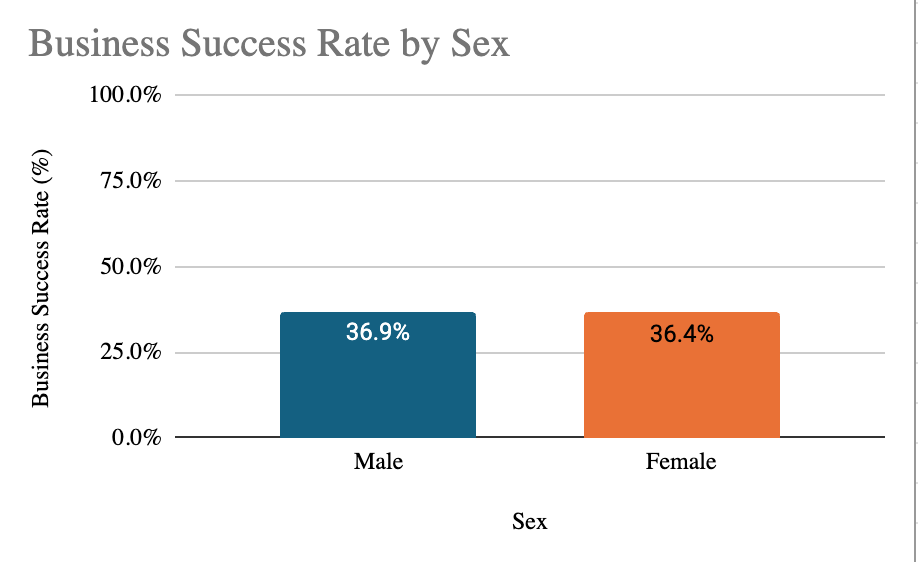}
{Business Success Rate by Sex}

\indent This near equality does not mean that gender is irrelevant to entrepreneurship. Chapter 4 shows a large difference in who enters business ownership, whereas Figure 11 conditions on respondents who established a business and have an observable outcome. Within that selected founder sample, however, the aggregate proxy-success rates are almost the same. The strongest descriptive contrast in the paper is therefore between a large gender difference at entry and little aggregate gender difference under the post-entry proxy. \\

\indent Educational attainment introduces more variation. Among high-school-educated founders, proxy success is 27.6\% for men (\(n=185\)) and 32.0\% for women (\(n=122\)). Among associate-degree holders, the rates are 45.5\% (\(n=44\)) and 37.5\% (\(n=40\)); among bachelor's-degree holders, 48.2\% (\(n=83\)) and 44.8\% (\(n=58\)); among master's-degree holders, 45.0\% (\(n=20\)) and 36.8\% (\(n=19\)); and among doctoral/professional founders, 62.5\% (\(n=16\)) and 22.2\% (\(n=9\)). \\

\researchfigure{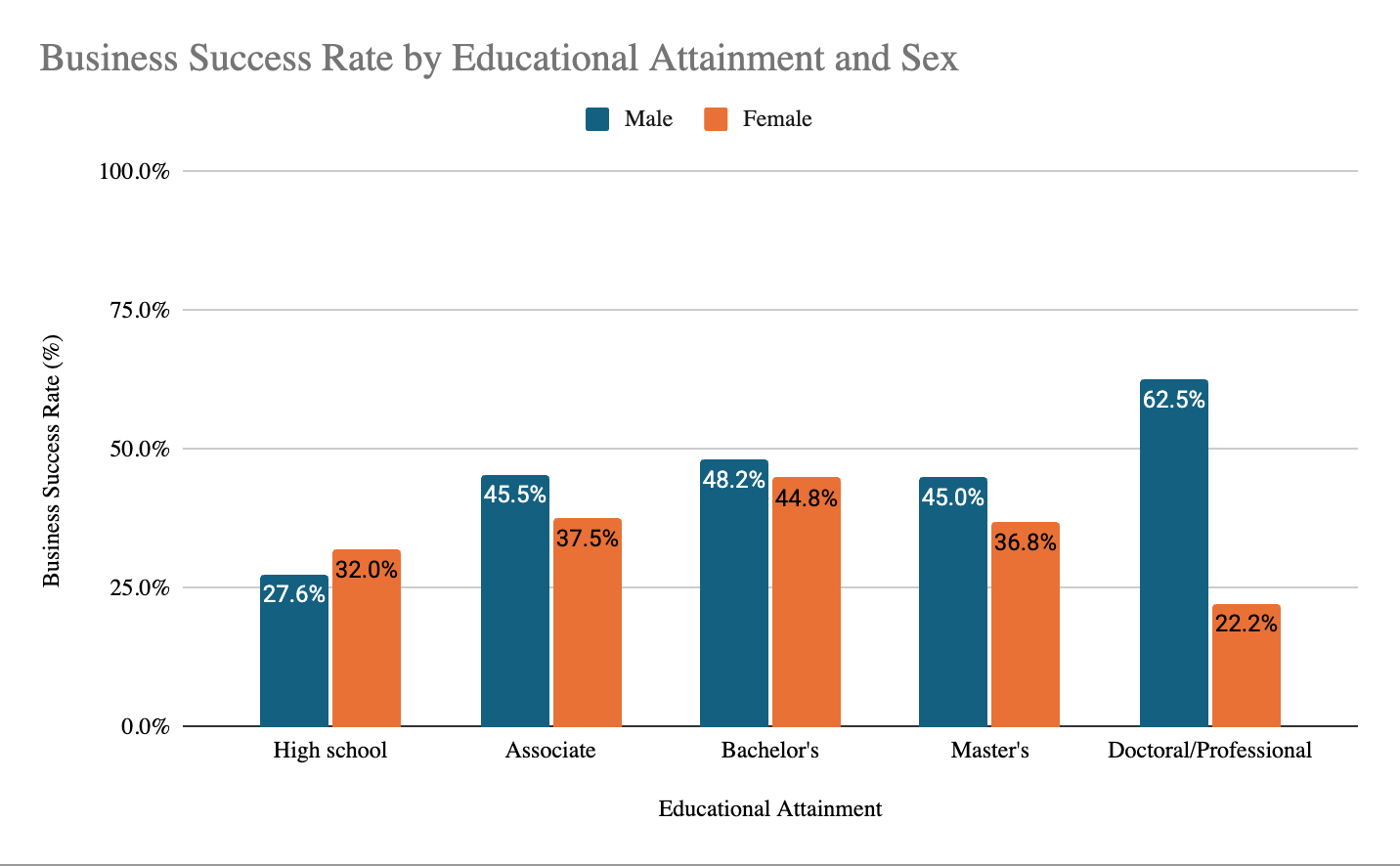}
{Business Success Rate by Educational Attainment and Sex}

\indent The more reliable part of Figure 12 is the contrast between high-school founders and the associate- or bachelor's-degree groups, where observed rates are higher for both sexes. Beyond that, the pattern is not monotonic and the samples become small. The 40.3-percentage-point doctoral/professional gender gap is particularly unstable because it is based on only 16 male and 9 female founders. It should not be interpreted as evidence of a general high-education gender effect. \\

\indent Maternal education also produces variation, but the pattern differs by sex. For male founders, proxy-success rates are 31.2\% (\(n=109\)) when the mother completed less than high school, 34.9\% (\(n=152\)) for high school, 36.4\% (\(n=44\)) for some college, 41.9\% (\(n=31\)) for college graduate, and 73.3\% (\(n=15\)) for postgraduate education. \\

\indent For female founders, the corresponding rates are 31.0\% (\(n=87\)), 36.8\% (\(n=106\)), 50.0\% (\(n=24\)), 46.2\% (\(n=26\)), and 28.6\% (\(n=7\)). \\

\researchfigure{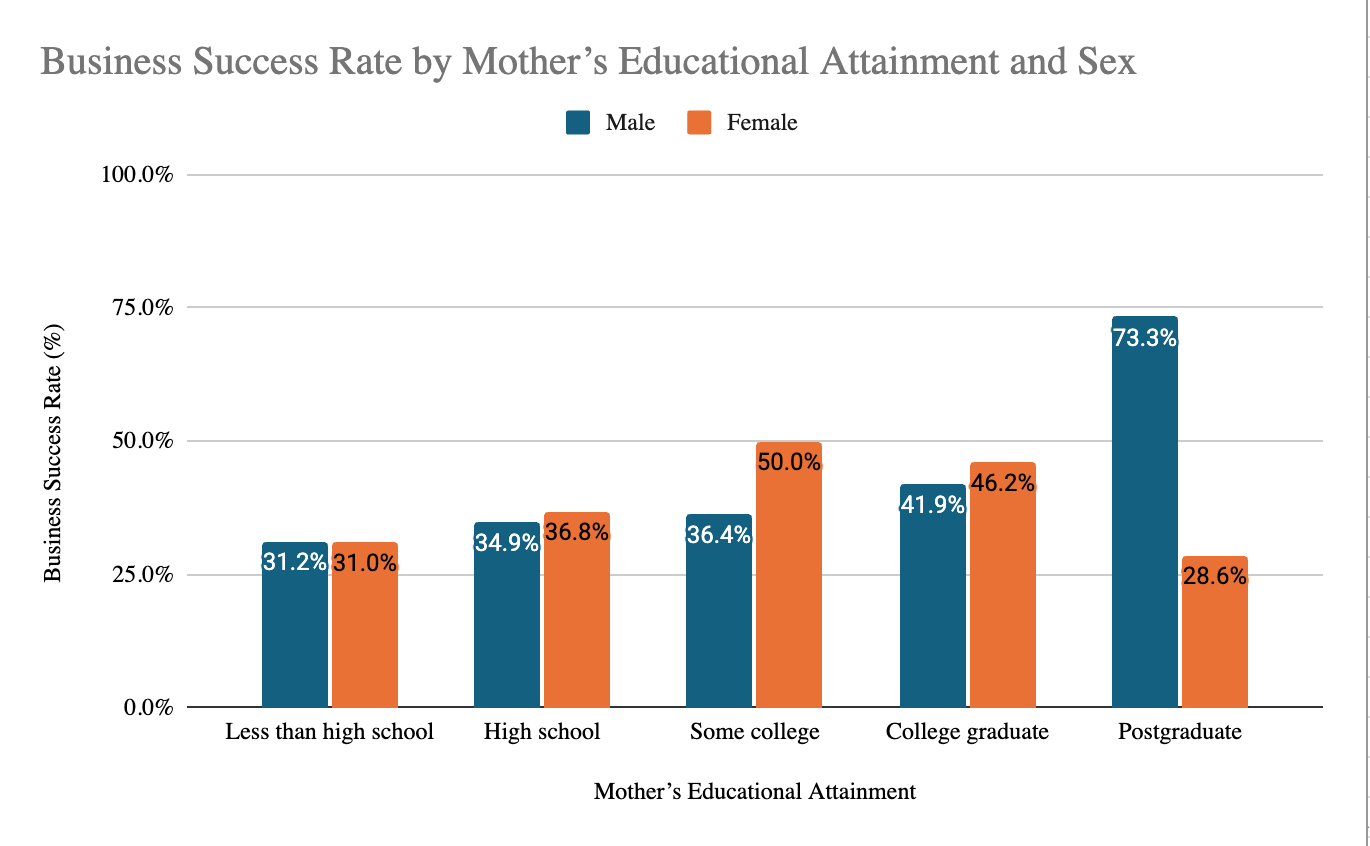}
{Business Success Rate by Mother's Educational Attainment and Sex}

\indent Across the larger lower and middle maternal-education groups, observed proxy success generally rises with maternal education, particularly among male founders. The female series rises through some college and then changes direction. The postgraduate comparison is not reliable as a broad pattern: the 73.3\% male rate is based on 15 founders and the 28.6\% female rate on only 7. Figure 13 therefore suggests possible background differences but does not support a simple monotonic maternal-education relationship for both sexes. \\

\indent Paternal education is even less consistent. When the father completed less than high school, proxy success is 32.4\% for men (\(n=102\)) and 29.4\% for women (\(n=68\)). The rates are 32.8\% (\(n=119\)) and 38.1\% (\(n=84\)) for high school; 46.2\% (\(n=39\)) and 26.9\% (\(n=26\)) for some college; 42.1\% (\(n=38\)) and 56.0\% (\(n=25\)) for college graduate; and 48.1\% (\(n=27\)) and 34.8\% (\(n=23\)) for postgraduate education. \\

\researchfigure{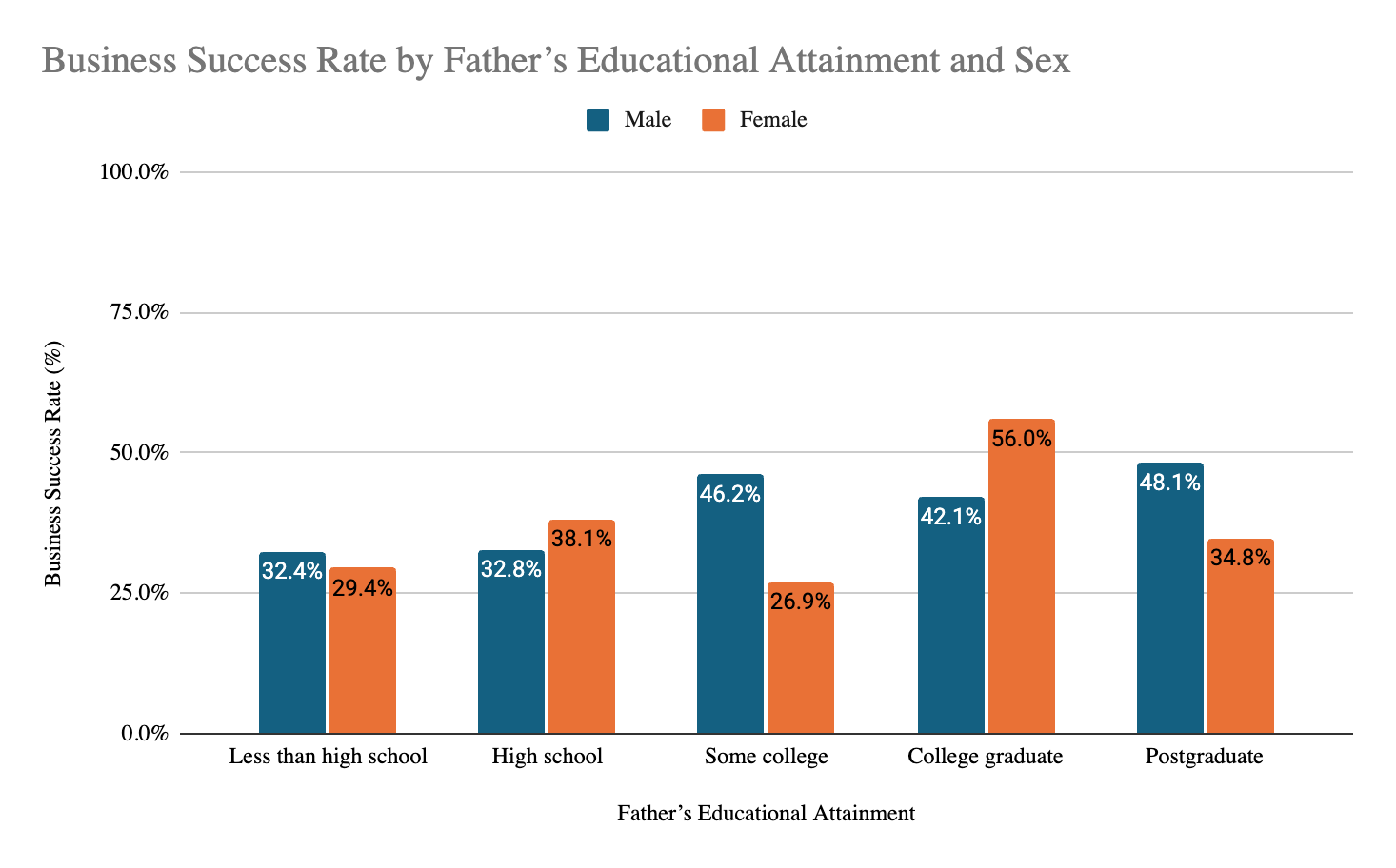}
{Business Success Rate by Father's Educational Attainment and Sex}

\indent The gender ordering reverses across paternal-education categories, and the rates do not form a common progression for men and women. The large differences in the some-college and college-graduate groups are also based on relatively small founder samples. Figure 14 therefore provides no clear descriptive evidence of a single paternal-education pattern in proxy success. \\

\indent Family income shows a different structure. Among male founders, proxy-success rates are 34.6\% in the low-income group (\(n=52\)), 36.7\% in the lower-middle group (\(n=60\)), 33.7\% in the upper-middle group (\(n=92\)), and 47.9\% in the high-income group (\(n=94\)). Among female founders, the rates are 28.2\% (\(n=39\)), 28.9\% (\(n=38\)), 42.6\% (\(n=54\)), and 40.6\% (\(n=64\)), respectively. \\

\researchfigure{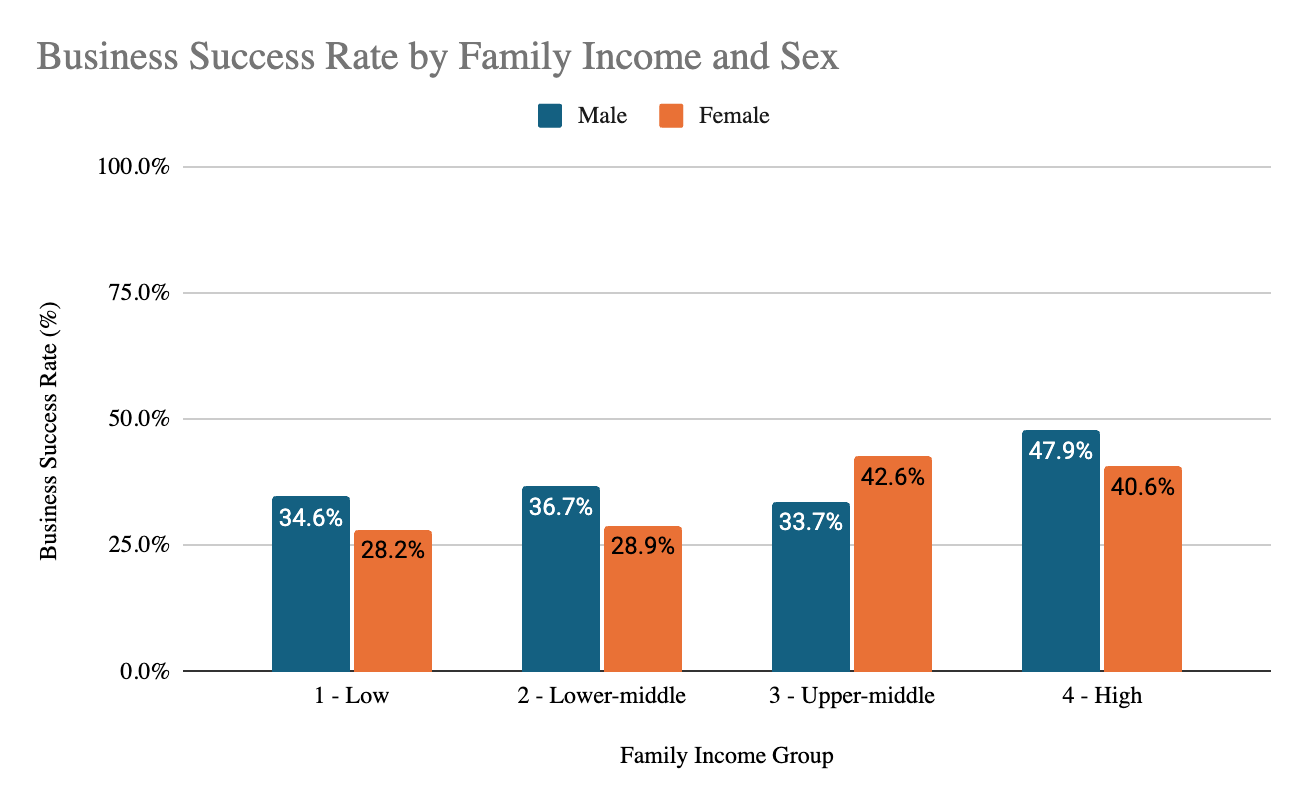}
{Business Success Rate by Family Income and Sex}

\indent The relationship is not monotonic for either sex. For men, the first three income groups cluster in the mid-30\% range, while the high-income group is higher at 47.9\%. For women, the upper-middle and high-income groups are both around 40\%, compared with about 28--29\% in the lower two groups. Thus, the figure is consistent with stronger observed outcomes in the upper part of the early-life family-income distribution, especially when the upper two female groups and the highest male group are compared with lower-income groups, but it does not support a smooth quartile-by-quartile gradient. \\

\indent Across Figures 11--15, the most stable gender result is the aggregate near equality in the proxy-success measure. Subgroup gender differences change direction across education, parental education, and income categories, and several of the largest differences occur where \(n\) is small. This is markedly different from Chapter 4, where the male ownership rate is higher in every corresponding background category. The descriptive evidence therefore indicates that the gender pattern is much more consistent for entrepreneurial entry than for the post-entry proxy among observed founders.


\section{Business Success and Nativity Status}

\indent Among founders with valid nativity and proxy-success information, the observed rate is 36.3\% for U.S.-born founders (\(n=576\)) and 45.9\% for foreign-born founders (\(n=37\)), a difference of approximately 9.6 percentage points. \\

\researchfigure{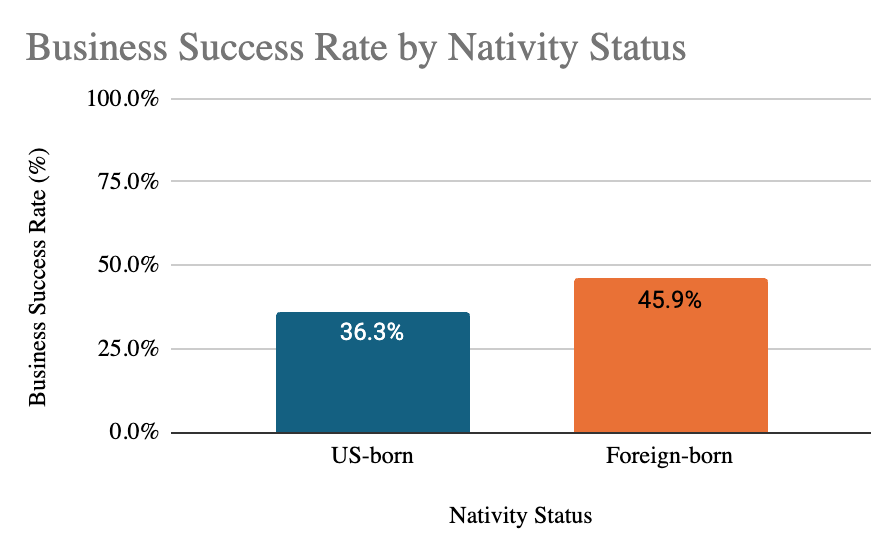}
{Business Success Rate by Nativity Status}

\indent The point estimate for foreign-born founders is higher, but the imbalance in sample size is substantial. Only 37 foreign-born founders contribute to the comparison. The 9.6-percentage-point difference should therefore be described as an observed sample difference rather than a stable foreign-born advantage. It does not establish that nativity causes differences in business outcomes. \\

\indent The sample-size problem becomes more severe once education is considered. Among high-school-educated founders, proxy success is 29.7\% for U.S.-born founders (\(n=279\)) and 27.8\% for foreign-born founders (\(n=18\)). Among associate-degree holders, the rates are 40.5\% (\(n=74\)) and 28.6\% (\(n=7\)); among bachelor's-degree holders, 44.9\% (\(n=136\)) and 100\% (\(n=5\)); and among master's-degree holders, 40.5\% (\(n=37\)) and 50.0\% (\(n=2\)). The U.S.-born doctoral/professional rate is 48.0\% (\(n=25\)), while no valid foreign-born estimate is available. \\

\researchfigure{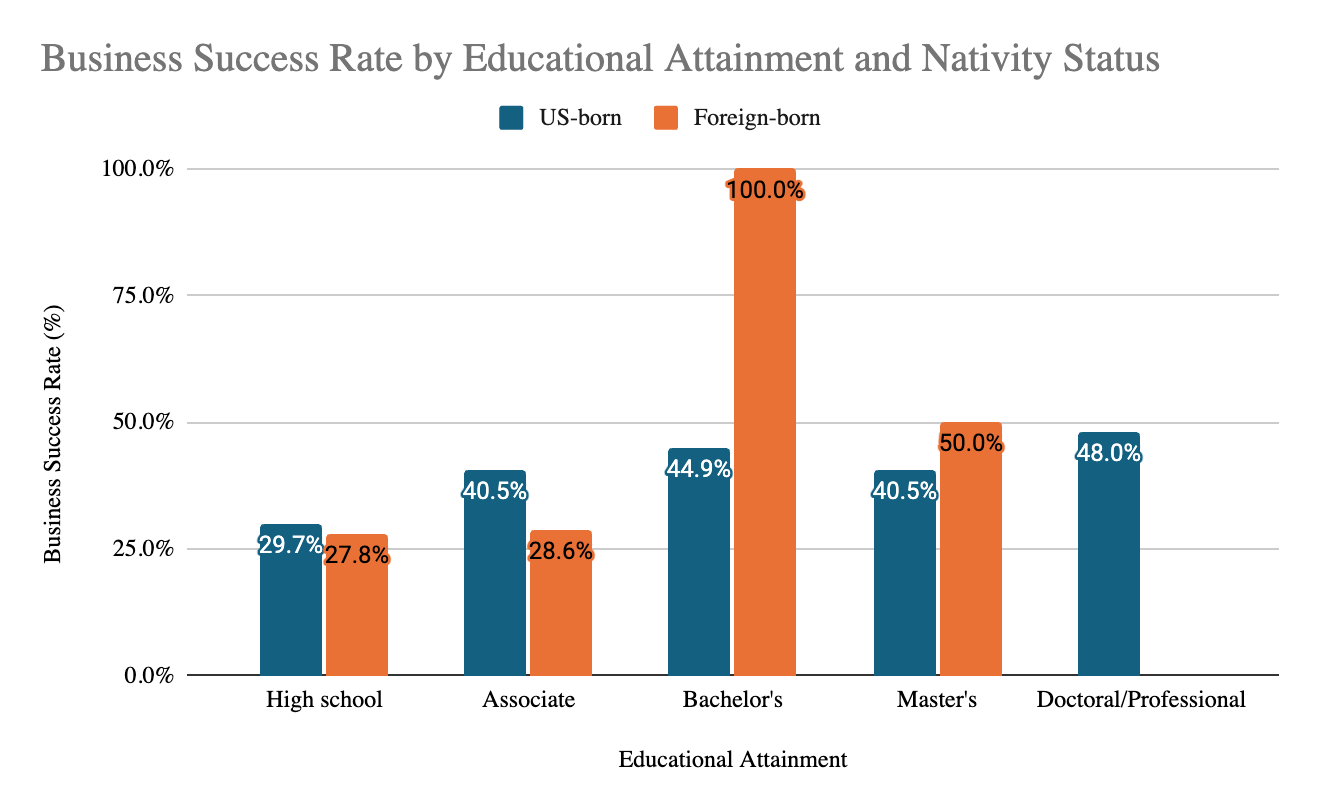}
{Business Success Rate by Educational Attainment and Nativity Status}

\indent The foreign-born percentages above the high-school level are based on seven or fewer founders. In particular, the 100\% bachelor's rate represents five observations and the 50.0\% master's rate represents two. These values describe the observed cases but cannot establish a stable educational pattern by nativity. The U.S.-born series shows higher rates in the associate, bachelor's, master's, and doctoral/professional groups than in the high-school group, but it is not monotonic across the higher degrees. \\

\indent Maternal education shows a clearer descriptive pattern only within the much larger U.S.-born founder sample. U.S.-born proxy-success rates are 30.2\% (\(n=159\)), 36.0\% (\(n=253\)), 40.0\% (\(n=65\)), 40.7\% (\(n=54\)), and 61.9\% (\(n=21\)) from less than high school through postgraduate education. \\

\indent The corresponding foreign-born rates are 40.0\% (\(n=25\)), 25.0\% (\(n=4\)), 100\% (\(n=1\)), 100\% (\(n=3\)), and 0\% (\(n=1\)). \\

\researchfigure{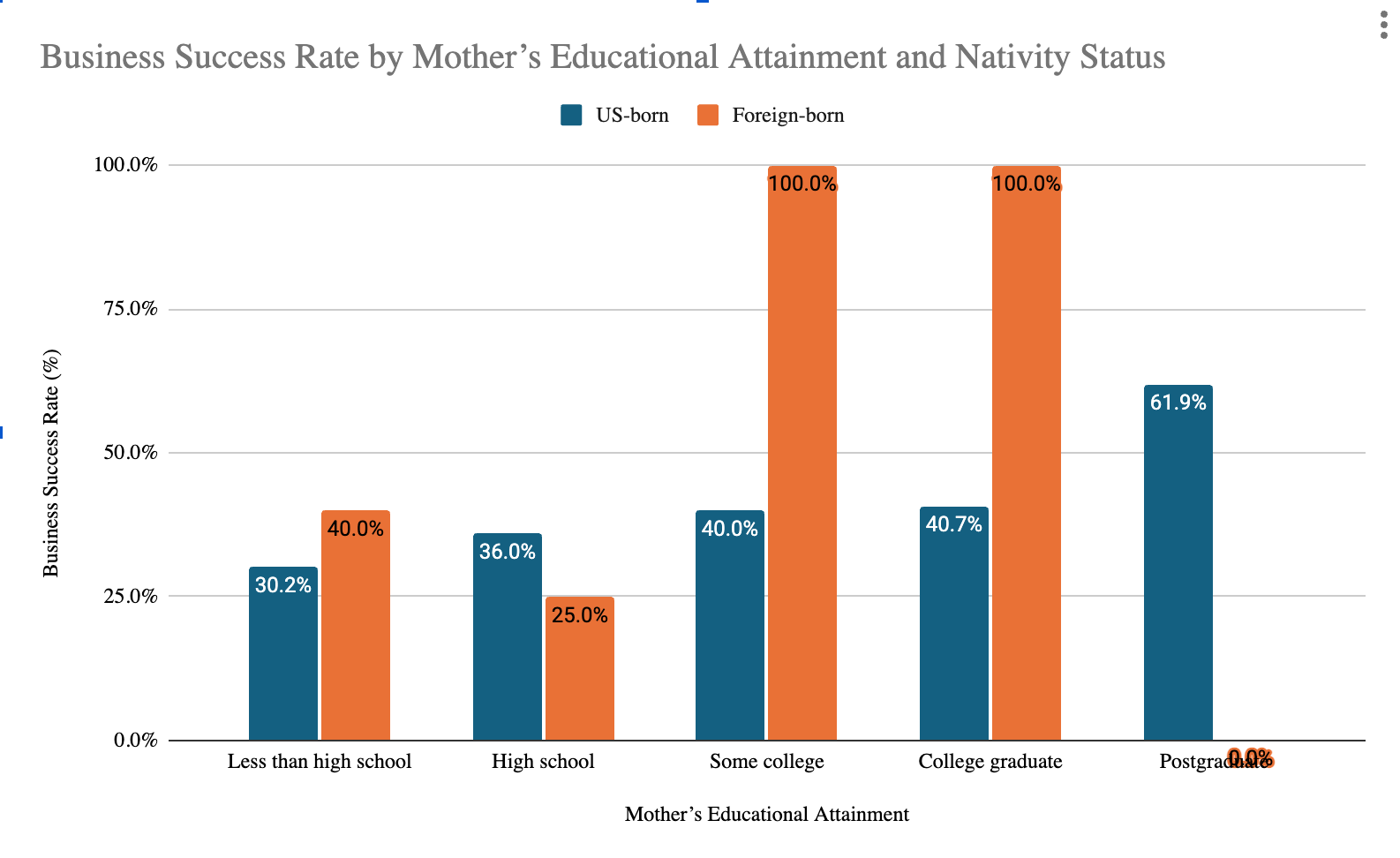}
{Business Success Rate by Mother's Educational Attainment and Nativity Status}

\indent For U.S.-born founders, Figure 18 shows a broadly increasing relationship between maternal education and the proxy outcome, with the highest rate in the postgraduate group. Even there, the postgraduate denominator is only 21, so the magnitude should be interpreted cautiously. The foreign-born series is not suitable for trend inference above the lowest category because four of the five groups contain four or fewer founders; its 0\% and 100\% values largely reflect individual cases rather than stable rates. \\

\indent Paternal education among U.S.-born founders shows a similarly broad but imperfect increase: 30.6\% (\(n=144\)), 34.9\% (\(n=192\)), 39.1\% (\(n=64\)), 45.9\% (\(n=61\)), and 40.4\% (\(n=47\)) across the five categories. Among foreign-born founders, the rates are 38.9\% (\(n=18\)), 28.6\% (\(n=7\)), 0\% (\(n=1\)), 100\% (\(n=2\)), and 66.7\% (\(n=3\)). \\

\researchfigure{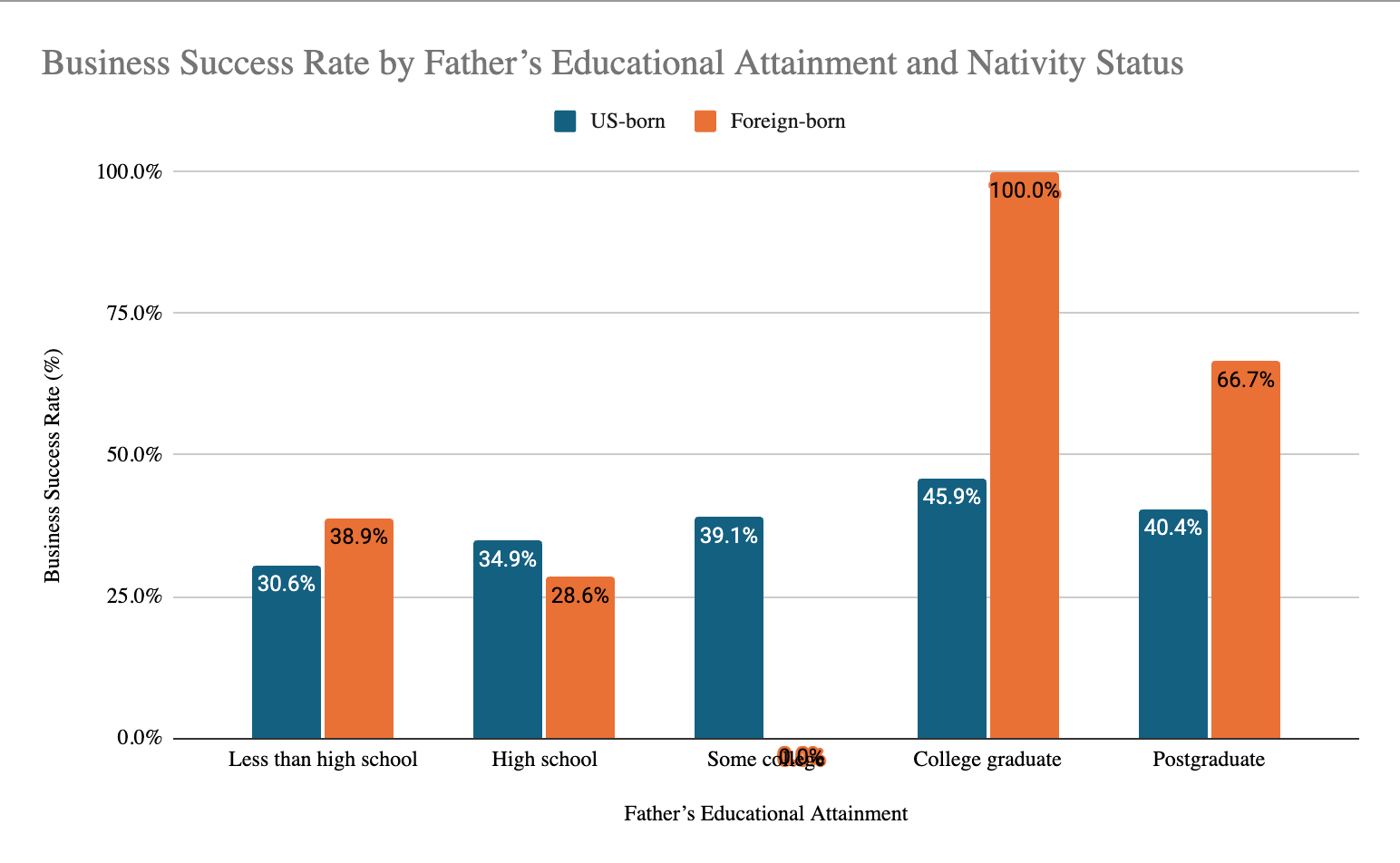}
{Business Success Rate by Father's Educational Attainment and Nativity Status}

\indent The U.S.-born pattern rises through the college-graduate category before declining in the postgraduate group, so it is better described as broadly increasing than monotonic. The foreign-born higher-education estimates are based on one to three founders and cannot support a meaningful paternal-education trend. \\

\indent Family income provides the most orderly background pattern within the U.S.-born founder sample. Proxy-success rates rise from 31.0\% in the low-income group (\(n=84\)) to 35.4\% in the lower-middle group (\(n=82\)), 35.6\% in the upper-middle group (\(n=135\)), and 44.8\% in the high-income group (\(n=154\)). For foreign-born founders, the corresponding rates are 60.0\% (\(n=5\)), 37.5\% (\(n=8\)), 57.1\% (\(n=7\)), and 50.0\% (\(n=4\)). \\

\researchfigure{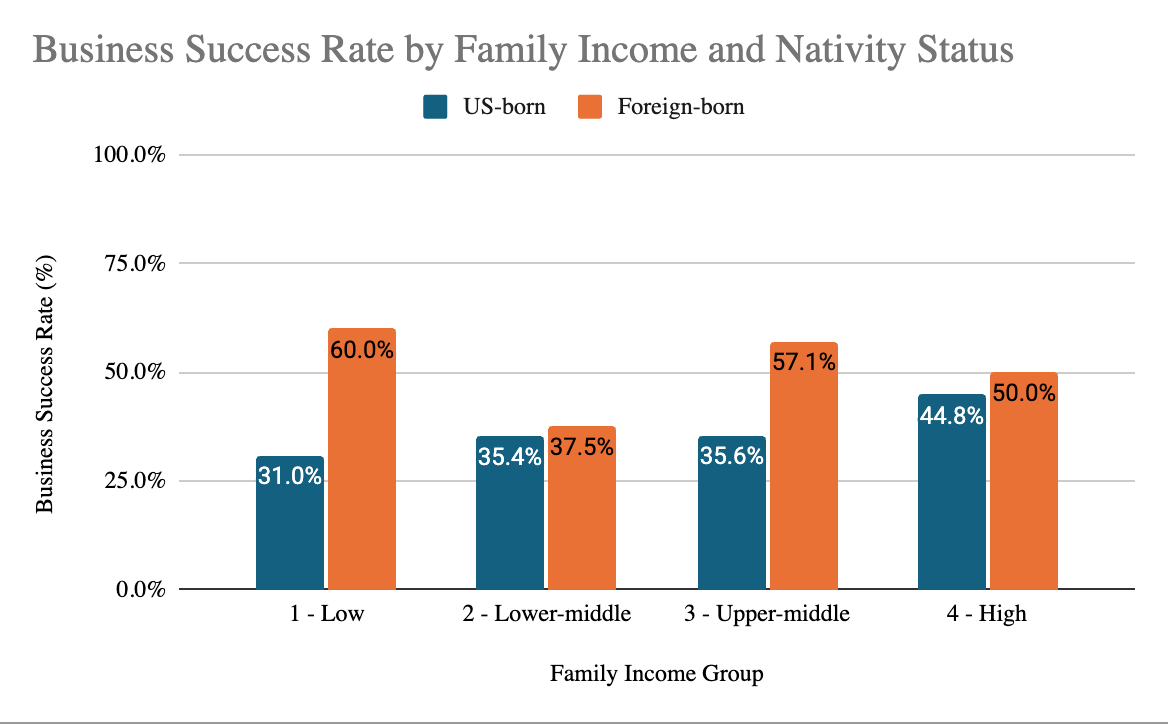}
{Business Success Rate by Family Income and Nativity Status}

\indent The U.S.-born series is consistent with higher observed proxy success at higher early-life family income, especially in the highest quartile. The foreign-born rates are higher than the U.S.-born rates in all four quartiles, but each foreign-born estimate is based on only four to eight founders. Those four comparisons therefore should not be interpreted as evidence that foreign-born founders systematically outperform U.S.-born founders at each income level. \\

\indent Overall, the aggregate nativity comparison in Figure 16 is the most interpretable result in this section, and even it is limited by the foreign-born \(n=37\). The subgroup figures mainly demonstrate how rapidly estimates become unstable when a small founder sample is divided by education or family background. The data support reporting a higher observed aggregate proxy-success rate among foreign-born founders, but they do not support a precise explanation of that difference or a stable subgroup pattern.

\clearpage


\chapter{CONCLUSION}
\vspace*{0.4cm}
\doublespacing

\indent This study separates entrepreneurship into two empirical stages: entry into business ownership and a post-entry outcome among founders. Using unweighted NLSY79 data, it compares descriptive rates across sex, nativity status, individual education, parental education, and early-life family income. The distinction between stages produces the paper's clearest result: the patterns associated with who enters entrepreneurship are more stable than the patterns observed in the smaller founder-outcome sample. \\

\indent The strongest entry finding concerns gender. Business ownership is reported by 18.2\% of male respondents (\(n=4{,}028\)) and 11.9\% of female respondents (\(n=4{,}173\)). More importantly, the male rate remains higher within every educational, maternal-education, paternal-education, and family-income category shown in Chapter 4. The size of the gap varies, and it becomes almost zero in the relatively small doctoral/professional group, but the ordering is otherwise consistent. These descriptive comparisons do not isolate a causal gender effect, yet they show that the aggregate difference is not confined to a single measured background category. \\

\indent Education and family background also structure entrepreneurial entry. Ownership tends to be higher in several more-educated groups, though the relationship is not monotonic. Maternal and paternal education are broadly positively associated with ownership, again with some reversals at the highest categories. Early-life family income shows a clearer split between the lower and upper halves of the distribution: for both men and women, ownership is higher in the upper-middle and high groups than in the two lower groups. These patterns are consistent with the broader literature on human capital and access to resources, but the present analysis cannot identify the mechanisms producing them. \\

\indent Nativity presents a weaker and less consistent entry pattern. The overall ownership rate is 15.1\% for U.S.-born respondents (\(n=7{,}455\)) and 13.9\% for foreign-born respondents (\(n=562\)). Across education, parental education, and income, the higher rate alternates between the two groups, and foreign-born denominators become small in several higher-education categories. The ownership analysis therefore does not provide evidence of a large, uniformly directed nativity gap comparable to the gender gap. \\

\indent The founder-outcome analysis changes the gender picture substantially. Under the study's business-success proxy, male founders have a rate of 36.9\% (\(n=371\)) and female founders 36.4\% (\(n=258\)). The aggregate difference is only about half a percentage point. Subgroup rates by education, parental education, and income vary, but the gender ordering frequently reverses and many large differences occur in small cells. The data therefore do not show the same consistent male advantage in the post-entry proxy that appears in business ownership. \\

\indent This contrast should be interpreted carefully. The founder analysis is conditional on having established a business and having a valid Business 01 outcome. Men and women who reach that stage may differ from the broader population in unobserved ways, and the outcome itself is only a proxy. The near-equal aggregate rates therefore do not prove that gender has no effect on business performance. They show a narrower result: among the observed founders in this analytical sample, the defined proxy outcome is nearly identical by sex even though business ownership is not. \\

\indent Education and family background produce less stable post-entry patterns than entry patterns. Associate- and bachelor's-degree founders often have higher observed proxy-success rates than high-school founders, but the highest educational categories are small and nonmonotonic. Maternal and paternal education show some upward tendencies in selected groups, especially among U.S.-born founders, yet the patterns are not uniform across sex. Early-life family income provides somewhat clearer evidence of stronger outcomes in the upper part of the distribution: male founders in the highest income group and female founders in the upper-middle and high groups have higher observed rates than their lower-income counterparts. Even these relationships remain descriptive and do not establish that early family income caused later business outcomes. \\

\indent Nativity in the founder analysis is especially constrained by sample size. Foreign-born founders have a higher observed aggregate proxy-success rate, 45.9\% (\(n=37\)), than U.S.-born founders, 36.3\% (\(n=576\)). Once foreign-born founders are divided by education, parental education, or income, however, many cells contain fewer than ten observations and several contain only one to three. The resulting 0\% and 100\% rates are mathematically correct descriptions of those cases but are not stable group estimates. The aggregate difference is therefore worth reporting, but its explanation remains unresolved. \\

\indent Several limitations define the scope of the conclusions. First, the analysis is descriptive and does not use regression adjustment or a causal identification strategy. Second, the business-success variable classifies continued ownership or sale as successful and shutdown as unsuccessful, but it does not measure profitability, valuation, growth, or financial return. Third, the analysis uses Business 01 rather than every business a respondent may have owned. Fourth, educational attainment may have been completed after business formation, while family income is measured much earlier in life. Fifth, the estimates are unweighted and should not automatically be treated as nationally representative. Finally, some founder subgroups, especially foreign-born groups, are too small for precise percentage comparisons. \\

\indent Within these limitations, the study supports one central conclusion: group differences in entrepreneurship look different depending on whether the focus is on entry or on outcomes after entry. In this sample, gender is strongly and consistently associated with business ownership, whereas the aggregate gender difference largely disappears under the post-entry proxy among observed founders. Nativity shows little consistent difference in ownership but a higher, much less precisely estimated foreign-born rate in the founder outcome. Education and family resources are associated with both stages in some comparisons, yet their post-entry patterns are less regular. Treating entrepreneurship as a single outcome would conceal these distinctions. Separating entry from post-entry outcomes therefore provides a more informative description of where observable group differences are concentrated.

\clearpage


\chapter*{REFERENCES}
\addcontentsline{toc}{chapter}{REFERENCES}
\vspace{1cm}

\singlespacing

\noindent Bates, Timothy. ``Entrepreneur Human Capital Inputs and Small Business Longevity.'' \textit{The Review of Economics and Statistics}, vol. 72, no. 4, 1990, pp. 551--559. \\

\vspace{0.4cm}

\noindent Dunn, Thomas, and Douglas Holtz-Eakin. ``Financial Capital, Human Capital, and the Transition to Self-Employment: Evidence from Intergenerational Links.'' \textit{Journal of Labor Economics}, vol. 18, no. 2, 2000, pp. 282--305. \\

\vspace{0.4cm}

\noindent Evans, David S., and Boyan Jovanovic. ``An Estimated Model of Entrepreneurial Choice under Liquidity Constraints.'' \textit{Journal of Political Economy}, vol. 97, no. 4, 1989, pp. 808--827. \\

\vspace{0.4cm}

\noindent Fairlie, Robert W., and Alicia M. Robb. ``Gender Differences in Business Performance: Evidence from the Characteristics of Business Owners Survey.'' \textit{Small Business Economics}, vol. 33, 2009, pp. 375--395. \\

\vspace{0.4cm}

\noindent Hout, Michael, and Harvey S. Rosen. ``Self-Employment, Family Background, and Race.'' \textit{Journal of Human Resources}, vol. 35, no. 4, 2000, pp. 670--692. \\

\vspace{0.4cm}

\noindent Hurst, Erik, and Annamaria Lusardi. ``Liquidity Constraints, Household Wealth, and Entrepreneurship.'' \textit{Journal of Political Economy}, vol. 112, no. 2, 2004, pp. 319--347. \\

\vspace{0.4cm}

\noindent Kerr, Sari Pekkala, and William R. Kerr. ``Immigrant Entrepreneurship in America: Evidence from the Survey of Business Owners 2007 \& 2012.'' \textit{Research Policy}, vol. 49, no. 3, 2020, article 103918. \\

\vspace{0.4cm}

\noindent Kim, Phillip H., Howard E. Aldrich, and Lisa A. Keister. ``Access (Not) Denied: The Impact of Financial, Human, and Cultural Capital on Entrepreneurial Entry in the United States.'' \textit{Small Business Economics}, vol. 27, 2006, pp. 5--22. \\

\vspace{0.4cm}

\noindent Lazear, Edward P. ``Entrepreneurship.'' \textit{Journal of Labor Economics}, vol. 23, no. 4, 2005, pp. 649--680. \\

\vspace{0.4cm}

\noindent National Longitudinal Surveys. ``Business Ownership: NLSY79 Topical Guide.'' \textit{National Longitudinal Survey of Youth 1979}, U.S. Bureau of Labor Statistics. \\

\vspace{0.4cm}

\noindent National Longitudinal Surveys. ``The NLSY79 Sample: An Introduction.'' \textit{National Longitudinal Survey of Youth 1979}, U.S. Bureau of Labor Statistics. \\

\end{document}